\documentclass[11pt]{amsart}
\usepackage{latexsym,amssymb,amsmath,amscd,amsthm}
\numberwithin{equation}{section}
\theoremstyle{remark}

\newtheorem{definition}{{\bf DEFINITION}}[section]

\newcommand{\bq}{\begin{equation}}
\newcommand{\bea}{\begin{array}}
\newcommand{\eea}{\end{array}}

\newcommand{\gD}{\Delta}
\newcommand{\gl}{\lambda}

\newcommand{\ot}{\otimes}
\newcommand{\mf}{\mathfrak}
\newcommand{\mc}{\mathcal}

\newcommand{\wg}{\wedge}

\newcommand{\ci}{\circ}

\newcommand{\go}{\omega}
\newcommand{\gO}{\Omega}
\newcommand{\gG}{\Gamma}
\newcommand{\gt}{\theta}

\newcommand{\gag}{\gamma}
\newcommand{\gd}{\delta}
\newcommand{\pp}{\partial}

\newcommand{\tl}{\tilde}
\newcommand{\na}{\nabla}
\newcommand{\gk}{\kappa}

\newcommand{\bs}{\blacksquare}

\newcommand{{\DDD}}{D\!\!\!\!\!\!-}

\newcommand{\bx}{\Box}
\newcommand{\mpt}{\mapsto}

\title{FISHER, K\"AHLER, WEYL, AND THE QUANTUM POTENTIAL}
\author{Robert Carroll\\University of Illinois, Urbana, IL 61801}

\date{July, 2004\thanks{email: rcarroll@math.uiuc.edu}}

\begin{document}

\bibliographystyle{plain}

\begin{abstract}
This is a basically expository article tracing connections of the quantum potential 
to Fisher information, to K\"ahler geometry of the projective Hilbert space of a quantum
system, and to the Weyl-Ricci scalar curvature of a Riemannian flat spacetime with quantum
matter.
\end{abstract}

\maketitle

\tableofcontents

\section{INTRODUCTION}
\renewcommand{\theequation}{1.\arabic{equation}}
\setcounter{equation}{0}

There is a comprehensive outline of quantum geometry in \cite{a12} (cf. also \cite{a15,
a16,a2,a3,a5,a6,b1,b34,c53,c2,c3,c4,c5,f1,g1,g2,h55,i1,j1,k50,k2,l1,m3,m2,m30,p3,p4,s1,
t1,w1}).  We will develop certain features and formulas in a ``hands on" approach following
\cite{a16,a2,b1,c53,c2,c3,c4,c5,p2,p3,p4,t1,w1} and spell out the nature of the K\"ahler
geometry for the projective Hilbert space of a quantum system along with the relation
between the Fisher metric and the Fubini-Study metric.  Then we go to
\cite{b2,c1,f2,g3,h5,h6,r1,r2} for discussion of connections between the quantum
potential and Fisher information.  Finally following \cite{c1,c16,s2} we indicate
connections of the quantum potential to the Weyl-Ricci scalar curvature of space time,
thus connecting quantum geometry, gravity, and Fisher information.  Relations of 
Fisher information to entropy are also sketched. Roughly the idea is that for H the
Hilbert space of a quantum system there is a natural quantum geometry on the
projective space $P(H)$ with inner product
${\bf (A1)}\,\,<\phi|\psi>=(1/2\hbar)g(\phi,\psi)+(i/2\hbar)\go(\phi,\psi)$ where
$g(\phi,\psi) =2\hbar\Re(\phi|\psi)$ is the natural Fubini-Study (FS) metric and
$g(\phi,\psi)=\go(\phi,J\psi)\,\,(J^2=-1)$.  On the other hand the FS metric is
proportional to the Fisher information metric of the form ${\bf
(A2)}\,\,Cos^{-1}|<\phi|\psi>|$.  Moreover (in 1-D for simplicity) ${\bf (A3)}\,\,{\mf
F}\propto
\int
\rho Qdx$ is a functional form of Fisher information where Q is the quantum potential
and
$\rho=|\psi|^2$.  Finally one can argue that in a Riemannian flat spacetime (with
quantum matter and Weyl geometry) the Weyl-Ricci scalar curvature is proportional to
Q.  We will explain this below and refer to \cite{c1} for more details and
perspective.

\section{QUANTUM GEOMETRY}
\renewcommand{\theequation}{2.\arabic{equation}}
\setcounter{equation}{0}

First we sketch the relevant symbolism for geometrical QM from
\cite{a12} without much philosophy; the philosophy is eloquently phrased 
in \cite{a12,a2,b34,c53,c2,k50,m30} for example.  Thus let H be the Hilbert space of QM
and write it as a real Hilbert space with a complex structure J.  The Hermitian inner
product is then
${\bf (B1)}\,\,<\phi,\psi>=(1/2\hbar)g(\phi,\psi)+(i/2\hbar)\go(\phi,\psi)$
(note $g(\phi,\psi)=2\hbar\Re(\phi,\psi)$ is the natural Fubini-Study (FS) metric and
this is discussed below - cf. \cite{c53,c2,c3,c4,c5}).  Here $g$ is a
positive definite real inner product and
$\go$ is a symplectic form (both strongly nondegenerate).  Moreover
${\bf (B2)}\,\,<\phi,J\psi>=i<\phi,\psi>$ and ${\bf
(B3)}\,\,g(\phi,\psi)=\go(\phi,J\psi)$.  Thus the triple $(J,g,\go)$ equips H with the
structure of a K\"ahler space.  
Now, from \cite{w2}, on a real
vector space V with complex structure J a Hermitian form satisfies $h(JX,JY)=h(X,Y)$.
Then V becomes a complex vector space via $(a+ib)X=aX+bJX$.  A Riemannian metric $g$
on a manifold M is Hermitian if $g(X,Y)=g(JX,JY)$ for $X,Y$ vector fields on M.  Let
$\na_X$ be he Levi-Civita connection for $g$ (i.e. parallel transport preserves inner
products and the torsion is zero - see \eqref{star} below).  A manifold M with J as
above is called almost complex.  A complex manifold is a paracompact Hausdorff space
with complex analytic patch transformation functions.  An almost complex M with
K\"ahler metric (i.e. $\na_XJ=0$) is called an almost K\"ahler manifold and if in
addition the Nijenhuis tensor vanishes it is a K\"ahler manifold (see \eqref{star}
below).  Here the defining equations for the Levi-Civita connection and the Nijenhuis
tensor are
\bq\label{star}
\gG^k_{ij}=\frac{1}{2}g^{hk}[\pp_ig_{jk}+\pp_jg_{ik}-\pp_kg_{ji}];\,\,
N(X,Y)=[JX,JY]-[X,Y]-J[X,JY]-J[JX,Y]
\end{equation}
Further discussion can be found in \cite{w2}.  Material on the Fubini-Study metric will
be provided later.
\\[3mm]\indent
Next (following \cite{a12}) by use of the canonical identification of the
tangent space (at any point of H) with H itself, $\gO$ is naturally extended to a
strongly nondegenerate, closed, differential 2-form on H, denoted also by $\gO$.  The
inverse of $\gO$ may be used to define Poisson brackets and Hamiltonian vector fields. 
Now in QM the observables may be viewed as vector fields, since linear operators
associate a vector to each element of the Hilbert space.  Moreover the  Schr\"odinger
equation, written here as
$\dot{\psi}=-(1/\hbar)J\hat{H}\psi$, motivates one to associate to each quantum
observable $\hat{F}$ the vector field ${\bf (B4)}\,\,
Y_{\hat{F}}(\psi)=-(1/\hbar)J\hat{F}\psi$.  The Schr\"odinger vector field is defined
so that the time evolution of the system corresponds to the flow along the
Schr\"odinger vector field and one can show that the vector field $Y_{\hat{F}}$, being
the generator of a one parameter family of unitary mappings on H, preserves both the
metric G and the symplectic form $\gO$.  Hence it is locally, and indeed globally,
Hamiltonian simce H is a linear space.  In fact the function which generates this
Hamiltonian vector field is simply the expectation value of $\hat{F}$.  To see this
write ${\bf (B5)}\,\,F:\,H\to{\bf R}$ via
$F(\psi)=<\psi,\hat{F}\psi>=<\hat{F}>=(1/2\hbar)G(\psi,\hat{F}\psi)$.  Now if $\eta$
is any tangent vector at $\psi$
\bq\label{x9}
(dF)(\eta)=\frac{d}{dt}<\psi+t\eta,\hat{F}(\psi+t\eta)>|_{t=0}=<\psi,\hat{F}\eta>+
<\eta,\hat{F}\psi>=
\end{equation}
$$=\frac{1}{\hbar}G(\hat{F}\psi,\eta)=\gO(Y_{\hat{F}},\eta)=
(i_{Y_{\hat{F}}}\gO)(\eta)$$
where one uses the selfadjointness of $\hat{F}$ and the definition of $Y_{\hat{F}}$
(recall the Hamiltonian vector field $X_f$ generated by f satisfies the equation
$i_{X_f}\gO=df$ and the Poisson bracket is defined via $\{f,g\}=\gO(X_f,X_g)$).
Thus the time evolution of any quantum mechanical system may be written in terms of
Hamilton's equation of classical mechanics; the Hamiltonian function is simply the
expectation value of the Hamiltonian operator.  Consequently Schr\"odinger's equation
is simply Hamilton's equation in disguise and for Poisson brackets we have
\bq\label{x10}
\{F,K\}_{\gO}=\gO(X_F,X_K)=\left<\frac{1}{i\hbar}[\hat{F},\hat{K}]\right>
\end{equation} 
where the right side involves the quantum Lie bracket.  Note this is not Dirac's 
correspondence principle since the Poisson bracket here is the quantum one determined
by the imaginary part of the Hermitian inner product.
Now look at the role played by G.  It enables one to define a real inner product
$G(X_F,X_K)$ between any two Hamiltonian vector fields and one expects that this
Riemann inner product is related to the Jordan product.  Indeed
\bq\label{x11}
\{F,K\}_{+}=\frac{\hbar}{2}G(X_F,X_K)=\left<\frac{1}{2}[\hat{F},\hat{K}]_{+}\right>
\end{equation}
Since the classical phase space is generally not equipped with a Riemannian metric
the Riemann product G does not have a classical analogue; however it does have a
physical interpretation.  One notes that the uncertainty of the observable $\hat{F}$ at
a state with unit norm is ${\bf
(B6)}\,\,(\gD\hat{F})^2=<\hat{F}^2>-<\hat{F}^2>=\{F,F\}_{+}-F^2$.  Hence the
uncertainty involves the Riemann bracket in a simple manner.  In fact Heisenberg's
uncertainty relation has a nice form as seen via
\bq\label{x12}
(\gD\hat{F})^2(\gD\hat{K})^2\geq\left<\frac{1}{2i}[\hat{F},\hat{K}]\right>^2+
\left<\frac{1}{2}[\hat{F}_{\perp},\hat{K}_{\perp}]_{+}\right>^2
\end{equation}
where $\hat{F}_{\perp}$ is the nonlinear operator defined by ${\bf
(B7)}\,\,\hat{F}_{\perp}(\psi)=\hat{F}(\psi)-F(\psi)$.
Thus $\hat{F}_{\perp}(\psi)$ is orthogonal to $\psi$ if $\|\psi\|=1$.  Using this one
can write \eqref{x12} in the form
\bq\label{x13}
(\gD\hat{F})^2(\gD\hat{K})^2\geq\left(\frac{\hbar}{2}\{F,K\}_{\gO}\right)^2+
(\{F,K\}_{+}-FK)^2
\end{equation}
The last expression in \eqref{x13} can be interpreted as the quantum covariance of
$\hat{F}$ and $\hat{K}$.
\\[3mm]\indent
The discussion in \cite{a12} continues in this spirit and is eminently worth reading;
however we digress here for a more ``hands on" approach following
\cite{c53,c2,c3,c4,c5}.  Assume H is separable with a complete orthonormal system
$\{u_n\}$ and for any $\psi\in H$ denote by $[\psi]$ the ray generated by $\psi$ while
$\eta_n=(u_n|\psi)$.  Define for $k\in{\bf N}$
\bq\label{3.3}
U_k=\{[\psi]\in P(H);\,\,\eta_k\ne 0\};\,\,\phi_k:\,U_k\to\ell^2({\bf C}):\,\,
\phi_k([\psi])=\left(\frac{\eta_1}{\eta_k},\cdots,\frac{\eta_{k-1}}{\eta_k},
\frac{\eta_{k+1}}{\eta_k},\cdots\right)
\end{equation}
where $\ell^2({\bf C})$ denotes square summable functions.  Evidently $P(H)=\cup_kU_k$
and $\phi_k\ci\phi_j^{-1}$ is biholomorphic.  It is easily shown that the structure is
independent of the choice of complete orthonormal system.  The coordinaes for $[\psi]$
relative to the chart $(U_k,\phi_k)$ are $\{z^k_n\}$ given via ${\bf (B8)}\,\,
z^k_n=(\eta_n/\eta_k)$ for $n<k$ and $z^k_n=(\eta_{n+1}/\eta_k)$ for $n\geq k$.
To convert this to a real manifold one can use $z^k_n=(1/\sqrt{2})(x^k_n+iy^k_n)$
with
\bq\label{3.4}
\frac{\pp}{\pp z^k_n}=\frac{1}{\sqrt{2}}\left(\frac{\pp}{\pp x^k_n}+i\frac{\pp}{\pp
y^k_n}\right);\,\,\frac{\pp}{\pp \bar{z}^k_n}=\frac{1}{\sqrt{2}}\left(\frac{\pp}{\pp
x_n^k}-i\frac{\pp}{\pp y^k_n}\right)
\end{equation}
etc.  Instead of nondegeneracy as a criterion for a symplectic form inducing a bundle
isomorphism between $TM$ and $T^*M$ one assumes here that a symplectic form on M is a
closed 2-form which induces at each point $p\in M$ a toplinear isomorphism between the
tangent and cotangent spaces at p.  For $P(H)$ one can do more than simply exhibit
such a natural symplectic form; in fact one shows that $P(H)$ is a K\"ahler manifold
(meaning that the fundamental 2-form is closed).  Thus one can choose a Hermitian
metric ${\bf (B9)}\,\,{\mf G}=\sum g^k_{mn}dz^k_m\ot d\bar{z}^k_n$ with
\bq\label{3.5}
g^k_{mn}=(1+\sum_i z^k_i\bar{z}^k_i)^{-1}\gd_{mn}-(1+\sum_1
z_i^k\bar{z}_i^k)^{-2}\bar{z}^k_mz^k_n
\end{equation}
relative to the chart $U_k,\phi_k)$.  The fundamental 2-form of the metric ${\mf G}$ is
${\bf (B10)}\,\,\go=i\sum_{m,n}g^k_{mn}dz_m^k\wg d\bar{z}_n^k$ and to show that this is
closed note that $\go=i\pp\bar{\pp}f$ where locally ${\bf (B11`)}\,\,f=log(1+\sum
z_i^k\bar{z}^k_i)$ (the local K\"ahler function).  Note here that $\pp+\bar{\pp}=d$
and $d^2=0$ implies $\pp^2=\bar{\pp}^2=0$ so $d\go=0$ and thus $P(H)$ is a K manifold.
\\[3mm]\indent 
Now on $P(H)$ the observables will be represented via a class
of real smooth functions on $P(H)$ (projective Hilbert space) called K\"ahlerian 
functions.  Consider a real smooth Banach manifold M with tangent space TM,
and cotangent space $T^*M$.  We remark that the extension of standard differential
geometry to the infinite dimensional situation of Banach manifolds etc. is essentially
routine modulo some functional analysis; there are a few surprises and some interesting
technical machinery but we omit all this here.  One should also use bundle terminology
at various places but we will not be pedantic about this.  One hopes here to simply
give a clear picture of what is happening.  Thus e.g. $L(T^*_xM,T_xM)$ denotes bounded
linear operators $T^*_xM\to T_xM$ and $L_n(T_xM,{\bf R})$ denotes bounded n-linear
forms on $T_xM$.  An almost complex structure is provided by a smooth section J of 
$L(TM)=$ vector bundle of bounded linear operators with fibres $L(T_xM)$ such that
$J^2=-1$.  Such a J is called integrable if its torsion is zero, i.e. $N(X,Y)=0$ with
N as in \eqref{star}.  An almost K\"ahler (K) manifold is a triple $(M,J,g)$ where M
is a real smooth Hilbert manifold, J is an almost complex structure, and $g$ is a K
metric, i.e. a Riemannian metric such that
\begin{itemize}
\item
$g$ is invariant; i.e. ${\bf (B12)}\,\,g_x(J_xX_x,J_xY_x)=g_x(X_x,Y_x)$.
\item
The fundamental two form of the metric is closed; i.e. ${\bf (B13)}\,\,\go_x(X_x,Y_x)=
g_x(J_xX_x,Y_x)$ is closed (which means $d\go=0$).
\end{itemize}
Note that an almost K manifold is canonically symplectic and if J is integrable one
says that M is a K manifold.  Now fix an almost K manifold $(M,J,g)$.  The form $\go$
and the K metric $g$ induce two top-linear isomorphisms $I_x$ and $G_x$ between 
$T_x^*M$ and $T_xM$ via ${\bf (B14)}\,\,\go_x(I_xa_x,X_x)=<a_x,X_x>$ and
$g_x(G_xa_z,X_x)=<a_x,X_x>$.  Denoting the smooth sections by $I,\,G$ one checks that
$G=J\ci I$.
\begin{definition}
For $f,h\in C^{\infty}(M,{\bf R})$ the Poisson and Riemann brackets are defined via
${\bf (B15)}\,\,\{f,h\}=<df,Idh>$ and ${\bf (B16)}\,\,((f,h))=<df,Gdh>$.  In view of
{\bf B14)} one can reformulate this as
\bq\label{3.6}
\{f,h\}=\go(Idf,Idh)=\go(Gdf,Gdh);\,\,((f,h))=g(Gdf,Gdh)=g(Idf,Idh)
\end{equation}
\end{definition}
\begin{definition}
For $f,h\in C^{\infty}(M,{\bf C})$ the K bracket is ${\bf
(B17)}\,\,<f,h>=((f,h))+i\{f,h\}$ and one defines products ${\bf (B18)}\,\,f\ci_{\nu}h=
(1/2)\nu((f,h))+fh$ ($\nu$ will be determined to be $\hbar$) and
$f*_{\nu}h=(1/2)\nu<f,h>+fh$. One observes also that
\bq\label{3.7}
f*_{\nu}h=f\ci_{\nu}h+(i/2)\nu\{f,h\};\,\,f\ci_{\nu}h=(1/2)(f*_{\nu}h+h*_{\nu}f);
\end{equation}
$$\{f,h\}=(1/i\nu)(f*_{\nu}h-h*_{\nu}f)$$
\end{definition}
\begin{definition}
For $f\in C^{\infty}(M,{\bf R})$ let $X=Idf$; then $f$ is called K\"ahlerian (K) if
${\bf (B19)}\,\,L_Xg=0$ where $L_X$ is the Lie derivative along X (recall
$L_Xf=Xf,\,\,L_XY=
[X,Y],\,\,L_X(\go(Y))=(L_X\go)(Y)+\go(L_X(Y)),\cdots$).  More generally if 
$f\in C^{\infty}(M,{\bf C})$ one says that $f$ is K if $\Re f$ and $\Im f$ are K;
the set of K functions is denoted by $K(M,{\bf R})$ or $K(M,{\bf C})$.
\end{definition}
\indent{\bf REMARK 2.1.}
In the language of symplectic manifolds $X=df$ is the Hamiltonian vector field
corresponding to $f$ and the condition $L_Xg=0$ means that the integral flow of X,
or the Hamiltonian flow of $f$, preserves the metric $g$.  From this follows also 
$L_XJ=0$ (since J is uniquely determined by $\go$ and $g$ via {\bf (B13)}).  Therefore
if $f$ is K the Hamiltonian flow of $f$ preserves the whole K structure.  Note also 
that $K(M,{\bf R})$ (resp. $K(M,{\bf C})$) is a Lie subalgebra of $C^{\infty}(M,{\bf
R})$ (resp. $C^{\infty}(M,{\bf C})$).$\hfill\bs$
\\[3mm]\indent
Now $P(H)$ is the set of one dimensional subspaces or rays of H; for every 
$x\in H/\{0\},\,\,[x]$ is the ray through $x$.  If H is the Hilbert space of a
Schr\"odinger quantum system then H represents the pure states of the system and 
$P(H)$ can be regarded as the state manifold (when provided with the differentiable
structure below).  One defines the K structure as follows.  On $P(H)$ one has an 
atlas $\{(V_h,b_h,C_h)\}$ where $h\in H$ with $\|h\|=1$.  Here $(V_h,b_h,C_h)$ is the
chart with domain $V_h$ and local model the complex Hilbert space $C_h$ where
\bq\label{3.8}
V_h=\{[x]\in P(H);\,(h|x)\ne 0\};\,\,C_h=[h]^{\perp};\,\,b_h:\,V_h\to C_h;\,\,[x]\to
b_h([x])=\frac{x}{(h|x)}-h
\end{equation}
This produces a analytic manifold structure on $P(H)$.  As a real manifold one uses
an atlas $\{(V_h,R\ci b_h,RC_h)\}$ where e.g. $RC_h$ is the realification of $C_h$
(the real Hilbert space with ${\bf R}$ instead of ${\bf C}$ as scalar field)
and $R:\,C_h\to RC_h;\,v\to Rv$ is the canonical bijection (note $Rv\ne\Re v$).
Now consider the form of the K metric relative to a chart $(V_h,R\ci b_h,RC_h)$ where
the metric $g$ is a smooth section of $L_2(TP(H),{\bf R})$ with local expression
$g^h:\,RC_h\to L_2(RC_h,{\bf R});\,Rz\mpt g^h_{Rz}$ where 
\bq\label{3.9}
g^h_{Rz}(Rv,Rw)=2\nu\Re\left(\frac{(v|w)}{1+\|z\|^2}-
\frac{(v|z)(z|w)}{(1+\|z\|^2)^2}\right)
\end{equation}
The fundamental form $\go$ is a section of $L_2(TP(H),{\bf R})$, i.e. 
$\go^h:\,RC_h\to L_2(RC_h,{\bf R});\,\,Rz\to \go^h_{Rz}$, given via
\bq\label{3.10}
\go^h_{Rz}(Rv,Rw)=2\nu\Im\left(\frac{(v|w)}{1+\|z\|^2}-
\frac{(v|z)(z|w)}{(1+\|z\|^2)^2}\right)
\end{equation}
\indent
Then using e.g. \eqref{3.9} for the FS metric in $P(H)$ consider
a Schr\"odinger Hilbert space with dynamics determined via ${\bf (B20)}\,\,{\bf
R}\times P(H)\to P(H):\,(t,[x])\mpt [exp(-(i/\hbar)tH)x]$ where H is a (typically
unbounded) self adjoint operator in H.  One thinks then of K\"ahler isomorphisms of
$P(H)$ (i.e. smooth diffeomorphisms $\Phi:\,P(H)\to P(H)$ with the properties
$\Phi^*J=J$ and $\Phi^*g=g$). If U is any unitary operator on H the map $[x]\mpt [Ux]$
is a K isomorphism of $P(H)$.  Conversely (cf. \cite{c3}) any K isomorphism of $P(H)$
is induced by a unitary operator U (unique up to phase factor).  Further for every self
adjoint operator A in H (possibly unbounded) the family of maps $(\Phi_t)_{t\in{\bf
R}}$ given via ${\bf (B21)}\,\,
\Phi_t:\,[x]\to [exp(-itA)x]$ is a continuous one parameter group of K isomorphisms of 
$P(H)$ and vice versa (every K isomorphism of $P(H)$ is induced by a self adjoint
operator where boundedness of A corresponds to smoothness of the $\Phi_t$).  Thus in the
present framework the dynamics of QM is described by a continuous one parameter group
of K isomorphisms, which automatically are symplectic isomorphisms (for the structure
defined by the fundamental form) and one has a Hamiltonian system.  Next ideally one can
suppose that every self adjoint operator represents an observable and these will be
shown to be in $1-1$ correspondence with the real K functions.
\begin{definition}
Let A be a bounded linear operator on H and denote by $<A>$ the mean value function
of A defined via ${\bf (B22)}\,\,<A>:\,P(H)\to {\bf C},\,\,[x]\mpt
<A>_{[x]}=(x|Ax)/\|x\|^2$.  The square dispersion is defined via ${\bf (B23)}\,\,
\gD^2A:\,P(H)\to{\bf C},\,\,[x]\mpt \gD^2_{[x]}A=<(A-<A>_{[x]})^2>_{[x]}$.
\end{definition}
These maps {\bf (B22)} and {\bf (B23)} are smooth and if A is self adjoint $<A>$ is
real,
$\gD^2A$ is nonnegative, and one can define $\gD A=\sqrt{\gD^2 A}$.  To obtain local
expressions one writes $<A>^h:\,C_h\to{\bf R}$ and $(d<A>)^h:\,C_h\to (C_h)^*$ via
${\bf (B24)}\,\,<A>^h(R)=(z+h)|A(z+h))/(1+\|z\|^2)$ and 
\bq\label{3.11}
<(d<A>)^h_{Rz}|Rv>=2\Re\left(\left.\frac{A(z+h)}{1+\|z\|^2}-\frac{(h|A(z+h))}{1+\|z\|^2}h-
\frac{(A(z+h)|z+h)}{(1+\|z\|^2)^2}z\right|Rv\right)
\end{equation}
Further the local expressions $X^h:\,RC_h\to RC_h$ and $Y^h:\,RC_h\to RC_h$ of the
vector fields $X=Id<A>$ and $Y=Gd<A>$ are 
\bq\label{3.12}
X^h(Rz)=(1/\nu)R(i(h|A(z+h))(z+h)-iA(z+h));
\end{equation}
$$Y^h(Rz)=(1/\nu)R(-(h|A(z+h))(z+h)+A(z+h))$$
One proves then (cf. \cite{c3,h5}) that the flow of the vector field $X=I d<A>$ is
complete and is given via ${\bf (B25)}\,\,\Phi_t([x])=[exp(-i(t/\nu)A)x]$.  This leads
to the statement that if $f$ is a complex valued function on $P(H)$ then $f$ is
K\"ahlerian if and only if there is a bounded operator A such that $f=<A>$ (cf.
Definition 3.2).  From the above it is clear that one should take $\nu=\hbar$ for QM if
we want to have
$<{\mf H}>$ represent Hamiltonian flow (${\mf H}\sim$ a Hamiltonian operator) and this
gives a geometrical interpretation of Planck's constant.  The following formulas are
obtained for the Poisson and Riemann brackets
\bq\label{3.13}
\{<A>,<B>\}^h(Rz)=\frac{(z+h|(1/i\nu)(AB-BA)(z+h))}{1+\|z\|^2};\,\,((<A>,<B>))^h(Rz)=
\end{equation}
$$\frac{1}{\nu}\frac{(z+h|(AB+BA)(z+h))}{1+\|z\|^2}-\frac{2}{\nu}\frac{(z+h|A(z+h))}
{1+\|z\|^2}\frac{(z+h|B(z+h))}{1+\|z\|^2}$$
This leads to the results
\begin{enumerate}
\item
$\{<A>,<B>\}=<(1/i\nu)[A,B]>$
\item
$((<A>,<B>))=(1/\nu)<AB+BA>-(2/\nu)<A><B>;\,\,((<A><A>))=(2/\nu)\gD^2A$
\item
$<<A>,<B>>=(2/\nu)(<AB>-<A><B>)$
\item
$<A>\ci_{\nu}<B>=(1/2)<AB+BA>$
\item
$<A>*_{\nu}<B>=<AB>$
\end{enumerate}
\indent
{\bf REMARK 2.2.}
One notes that (setting $\nu=\hbar$) item 1 gives the relation between Poisson brackets
and commutators in QM.  Further the Riemann bracket is the operation needed to compute
the dispersion of observables.  In particular putting $\nu=\hbar$ in item 2 one sees
that for every observable $f\in K(P(H),{\bf R})$ and every state $[x]\in P(H)$ the
results of a large number of measurements of $f$ in the state $[x]$ are distributed
with standard deviation $\sqrt{\hbar/2)((f,f))([x])}$ around the mean value $f([x])$.
This explains the role of the Riemann structure in QM, namely it is the structure 
needed for the probabilistic description of QM.  Moreover the $\ci_{\nu}$ product
corresponds to the Jordan product between operators (cf. item 5) and item 4 tells us
that the $*_{\nu}$ product corresponds to the operator product.  This allows one to
formulate a functional representation for the algebra $L(H)$.  Thus put
${\bf (B26)}\,\,\|f\|_{\nu}=\sqrt{sup_{[x]}(\bar{f}*_{\nu}f)([x])}$.  Equipped with
this norm $K(P(H),{\bf C})$ becomes a $W^*$ algebra and the map of $W^*$ algebras
between $K(P(H),{\bf C}$ and $L(H)$ is an isomorphism.  This makes it possible to
develop a general functional representation theory for $C^*$ algebras generalizing the
classical spectral representation for commutative $C^*$ algebras.  The K manifold
$P(H)$is replaced by a topological fibre bundle in which every fibre is a K manifold
isomorphic to a projective space.  In particular a nonzero vector $x\in H$ is an
eigenvector of A if and only if $d_{[x]}<A>=0$ or equivalently if and only if $[x]$ is
a fixed point for the vector field $Id<A>$ (in which case the corresponding eigenvalue
is $<A>_{[x]}$).$\hfill\bs$
\\[3mm]\indent

\section{PROBABILITY ASPECTS}
\renewcommand{\theequation}{3.\arabic{equation}}
\setcounter{equation}{0}

We go here to \cite{a12,a16,a2,b1,b34,c53,c3,c4,f1,g1,g2,h5,h6,l1,m1,p2,p3,p4,r1,t1,w1};
some of this will be somewhat disjointed but we will organize it later.  First from
\cite{b1,w1} one defines a (Riemann) metric (statistical distance) on the space of
probability distributions ${\mc P}$ of the form ${\bf (C1)}\,\,ds^2_{PD}=
\sum(dp_j^2/p_j)=\sum p_j(dlog(p_j))^2$.  Here one thinks of the central limit theorem
and a distance between probability distributions distinguished via a Gaussian
$exp[-(N/2)(\tl{p}_j-p_j)^2/p_j]$ for two nearby distributions (involving N samples
with probabilities $p_j,\,\tl{p}_j$).  This can be generalized to quantum mechanical
pure states via (note $\psi\sim \sqrt{p}exp(i\phi)$ in a generic manner)
\bq\label{4.1}
|\psi>=\sum\sqrt{p_j}e^{i\phi_j}|j>;\,\,|\tl{\psi}>=|\psi>+|d\psi>=\sum \sqrt{p_j+dp_j}
e^{i(\phi_j+d\phi_j)}|j>
\end{equation}
Normalization requires $\Re(<\psi|d\psi>)=-1/2<d\psi|d\psi>$ and measurements described
by the one dimensional projectors $|j><j|$ can distinguish $|\psi>$ and $|\tl{\psi}>$
according to the metric {\bf (C1)}.  The maximum (for optimal disatinguishability) is
given by the Hilbert space angle ${\bf (C2)}\,\,cos^{-1}(|<\tl{\psi}|\psi>|)$ and the
corresponding line element ($PS\sim$ pure state)
\bq\label{4.2}
\frac{1}{4}ds^2_{PS}=[cos^{-1}(|<\tl{\psi}|\psi>|)]^2\sim 1-|<\tl{\psi}|\psi>|^2=
<d\psi_{\perp}|d\psi_{\perp}>\sim
\end{equation}
$$\sim \frac{1}{4}\sum\frac{dp_j^2}{p_j}+\left[\sum p_jd\phi_j^2-(\sum
p_jd\phi_j)^2\right]$$
(called the Fubini-Study (FS) metric) is the natural metric on the manifold of Hilbert
space rays.  Here ${\bf (C3)}\,\,|d\psi_{\perp}>=|d\psi>-|\psi><\psi|d\psi>$ is the
projection of
$|d\psi>$ orthogonal to $|\psi>$.  Note that if $cos^{-1}(|<\tl{\psi}|\psi>|=\gt$ then 
$cos(\gt)=|<\tl{\psi}|\psi>|$ and $cos^2(\gt)=|<\tl{\psi}|\psi>|^2=1-Sin^2(\gt)\sim
1-\gt^2$ for small $\gt$.  Hence $\gt^2\sim 1-cos^2(\gt)=1-|<\tl{\psi}|\psi>|^2$.
The term in square brackets (the
variance of phase changes) is nonnegative and an appropriate choice of basis makes it
zero.  In
\cite{b1} one then goes on to discuss distance formulas in terms of density operators
and Fisher information but we omit this here.  Generally as in \cite{w1} one observes
that the angle in Hilbert space is the only Riemannian metric on the set of rays which
is invariant uder unitary transformations.
In any event ${\bf (C4)}\,\,ds^2=\sum(dp_i^2/p_i),\,\,\sum p_i=1$ is referred to as the
Fisher metric (cf. \cite{m1}).  Note in terms of $dp_i=\tl{p}_i-p_i$ one can write
$d\sqrt{p}=(1/2)dp/\sqrt{p}$ with $(d\sqrt{p})^2=(1/4)(dp^2/p)$ and think of 
$\sum(d\sqrt{p_i})$ as a metric.  Alternatively from $cos^{-1}(|<\tl{\psi}|\psi>|$ one
obtains ${\bf (C5)}\,\,ds_{12}=cos^{-1}(\sum \sqrt{p_{1i}}\sqrt{p_{2i}})$ as a distance
in ${\mc P}$.  Note from {\bf (C3)} that ${\bf
(C6)}\,\,ds_{12}^2=4cos^{-1}|<\psi_1|\psi_2>|\sim 4(1-|(\psi_1|\psi_2)|^2\equiv
4(<d\psi|d\psi>-<d\psi|\psi><\psi|d\psi>)$ begins to look like a FS metric before passing
to projective coordinates.  In this direction we observe from \cite{m1} that the FS
metric as in \eqref{3.5} can be expressed also via 
\bq\label{4.3}
\pp\bar{\pp}log(|z|^2)=\phi=\frac{1}{|z|^2}\sum dz_i\wg d\bar{z}_i-\frac{1}{|z|^4}
\left(\sum \bar{z}_idz_i\right)\wg\left(\sum z_id\bar{z}_i\right)
\end{equation}
so for $v\sim \sum v_i\pp_i+\bar{v}_i\bar{\pp}_i$ and $w\sim \sum
w_i\pp_i+\bar{w}_i\bar{\pp}_i$ and $|z|^2=1$ one has ${\bf
(C7)}\,\,\phi(v,w)=(v|w)-(v|z)(z|w)$ (cf. \eqref{3.9}).

\subsection{FISHER INFORMATION}

We summarized in \cite{c1} various results on Fisher information, entropy, and the
Schr\"odinger equation (SE) followig \cite{f2,f3,g3,h5,h6,j2,r1,r2}.  Thus
first recall that the
classical Fisher information associated with translations of a 1-D observable X with
probability density
$P(x)$ is 
\bq\label{4.4}
F_X=\int dx\,P(x)([log(P(x)]')^2>0
\end{equation}
One has a well known Cramer-Rao inequality ${\bf (C8)}\,\,Var(X)\geq F_X^{-1}$
where $Var(X)\sim$ variance of X.  A Fisher length for X is defined via ${\bf
(C9)}\,\,\gd X=F_X^{-1/2}$ and this quantifies the length scale over which $p(x)$
(or better $log(p(x))$) varies appreciably.  Then the root mean square deviation
$\gD X$ satisfies ${\bf (C9)}\,\,\gD X\geq \gd X$.  Let now 
P be the momentum observable conjugate to X, and $P_{cl}$ a classical
momentum observable corresponding to the state $\psi$ given via ${\bf (C10)}\,\,
p_{cl}(x)=(\hbar/2i)[(\psi'/\psi)-(\bar{\psi}'/\bar{\psi})]$.   
One has the
identity ${\bf (C11)}\,\,<p>_{\psi}=<p_{cl}>_{\psi}$ following from {\bf (C10)}
with integration by parts. 
Now define the nonclassical momentum by $p_{nc}=p-p_{cl}$ 
and one shows then ${\bf (C12)}\,\,
\gD X\gD p\geq \gd X\gD p\geq \gd X\gD p_{nc}=\hbar/2$.
Now go to \cite{h6} now where two proofs are given for the derivation of the SE from
the exact uncertainty principle (as in {\bf (C12)} - cf. \cite{h5,h6}).  Thus consider a
classical ensemble of n-dimensional particles of mass m moving under a potential V.  The
motion can be described via the HJ and continuity equations
\bq\label{4.5}
\frac{\pp s}{\pp t}+\frac{1}{2m}|\na s|^2+V=0;\,\,\frac{\pp P}{\pp t}+
\na\cdot\left[P\frac{\na s}{m}\right]=0
\end{equation}
for the momentum potential $s$ and the position probability density P
(note that we have interchanged p and P from \cite{h6} - note also there is no quantum
potential and this will be supplied by the information term).   These
equations follow from the variational principle $\gd L=0$ with Lagrangian
${\bf (C13)}\,\,
L=\int dt\,d^nx\,P\left[(\pp s/\pp t)+(1/2m)|\na s|^2+V\right]$.
It is now assumed that the classical Lagrangian must be modified due to the existence
of random momentum fluctuations.  The nature of such fluctuations is immaterial for
(cf. \cite{h6} for discussion) and one can assume that the momentum associated with
position x is given by ${\bf (C14)}\,\,p=\na s + N$ where the fluctuation term N
vanishes on average at each point x.  Thus s changes to being an average momentum
potential.  It follows that the average kinetic energy $<|\na s|^2>/2m$ appearing in
{\bf (C13)} should be replaced by $<|\na s+N|^2>/2m$ giving rise to
\bq\label{4.6}
L'=L+(2m)^{-1}\int dt<N\cdot N>=L+(2m)^{-1}\int dt(\gD N)^2
\end{equation}
where $\gD N=<N\cdot N>^{1/2}$ is a measure of the strength of the quantum fluctuations. 
The additional term is specified uniquely, up to a multiplicative constant, by the
following three assumptions, namely
\begin{enumerate}
\item
Action principle:  $L'$ is a scalar Lagrangian with respect to the fields P and s
where the principle $\gd L'=0$ yields causal equations of motion.  Thus for some
scalar function $f$ one has $(\gD N)^2=\int d^nx\,pf(P,\na P,\pp P/\pp t,s,\na s,\pp s/\pp
t,x,t)$.
\item
Additivity:  If the system comprises two independent noninteracting subsystems with 
$P=P_1P_2$ then the Lagrangian decomposes into additive subsystem contributions; thus
$f=f_1+f_2$ for $P=P_1P_2$.
\item
Exact uncertainty:  The strength of the momentum fluctuation at any given time is
determined by and scales inversely with the uncertainty in position at that time.  
Thus $\gD N\to k\gD N$ for $x\to x/k$.  Moreover since position
uncertainty is entirely characterized by the probability density P at any given time
the function $f$ cannot depend on $s$, nor explicitly on $t$, nor on $\pp P/\pp t$.
\end{enumerate}
This leads to the result that ${\bf (C15)}\,\,(\gD N)^2=c\int d^nx\,P|\na
log(P)|^2$ where c is a positive universal constant (cf. \cite{h6}).  Further for
$\hbar=2\sqrt{c}$ and $\psi=\sqrt{P}exp(is/\hbar)$ the equations of motion for p and s
arising from $\gd L'=0$ are ${\bf (C16)}\,\,
i\hbar\frac{\pp\psi}{\pp t}=-\frac{\hbar^2}{2m}\na^2\psi+V\psi$.
\\[3mm]\indent
{\bf REMARK 3.1.}
In order to relate this to Fisher information we sketch here for simplicity and clarity
another derivation of the SE along similar ideas following
\cite{r1}.  Let
$P(y^i)$ be a probability density and $P(y^i+\gD y^i)$ be the density resulting from a
small change in the $y^i$.  Calculate the cross entropy via
\bq\label{4.7}
J(P(y^i+\gD y^i):P(y^i))=\int P(y^i+\gD y^i)log\frac{P(y^i+\gD y^i)}{P(y^i)}d^ny\simeq
\end{equation}
$$\simeq\left[\frac{1}{2}\int \frac{1}{P(y^i)}\frac{\pp P(y^i)}{\pp y^i}\frac
{\pp P(y^i)}{\pp y^k)}d^ny\right]\gD y^i\gD y^k=I_{jk}\gD y^i\gD y^k$$
The $I_{jk}$ are the elements of the Fisher information matrix.  The most general
expression has the form
\bq\label{4.8}
I_{jk}(\gt^i)=\frac{1}{2}\int\frac{1}{P(x^i|\gt^i)}\frac{\pp P(x^i|\gt^i)}{\pp \gt^j}
\frac{\pp P(x^i|\gt^i)}{\pp \gt^k}d^nx
\end{equation}
where $P(x^i|\gt^i)$ is a probability distribution depending on parameters $\gt^i$ in
addition to the $x^i$.  For ${\bf (C17)}\,\,P(x^i|\gt^i)=P(x^i+\gt^i)$ one recovers
\eqref{4.7} (straightforward - cf. \cite{r1}).  If P is defined over an n-dimensional
manifold with positive inverse metric $g^{ik}$ one obtains a natural definition of the
information associated with P via
\bq\label{4.9}
I=g^{ik}I_{ik}=\frac{g^{ik}}{2}\int\frac{1}{P}\frac{\pp P}{\pp y^i}\frac{\pp P}{\pp
y^k}d^ny
\end{equation}
Now in the HJ formulation of classical mechanics the equation of motion takes the form
\bq\label{4.10}
\frac{\pp S}{\pp t}+\frac{1}{2}g^{\mu\nu}\frac{\pp S}{\pp x^{\mu}}\frac{\pp S}
{\pp x^{\nu}}+V=0
\end{equation}
where $g^{\mu\nu}=diag(1/m,\cdots,1/m)$.  The velocity field $u^{\mu}$ is given by
${\bf C18)}\,\,u^{\mu}=g^{\mu\nu}(\pp S/\pp x^{\nu})$.  When the exact coordinates
are unknown one can describe the system by means of a probability density
$P(t,x^{\mu}$ with $int Pd^nx=1$ and ${\bf (C19)}\,\,
(\pp P/\pp t)+(\pp/\pp x^{\mu})(Pg^{\mu\nu}(\pp S/\pp x^{\nu})=0$.  These equations
completely describe the motion and can be derived from the Lagrangian
${\bf (C20)}\,\,
L_{CL}=\int P\left\{(\pp S/\pp t)+(1/2)g^{\mu\nu}(\pp S/\pp x^{\mu})
(\pp S/\pp x^{\nu})+V\right\}dtd^nx$
using fixed endpoint variation in S and P.  Quantization is obtained by adding a term
proportional to the information I defined in \eqref{4.9}.  This leads to
\bq\label{4.11}
L_{QM}=L_{CL}+\gl I=\int P\left\{\frac{\pp S}{\pp t}+\frac{1}{2}g^{\mu\nu}\left[
\frac{\pp S}{\pp x^{\mu}}\frac{\pp S}{\pp x^{\nu}}+\frac{\gl}{P^2}\frac{\pp P}
{\pp x^{\mu}}\frac{\pp P}{\pp x^{\nu}}\right]+V\right\}dtd^nx
\end{equation}
Fixed endpoint variation in S leads again to {\bf (C19)} while variation in P leads to
\bq\label{4.12}
\frac{\pp S}{\pp t}+\frac{1}{2}g^{\mu\nu}\left[\frac{\pp S}{\pp x^{\mu}}\frac{\pp S}
{\pp x^{\nu}}+\gl\left(\frac{1}{P^2}\frac{\pp P}{\pp x^{\mu}}\frac{\pp P}{\pp x^{\nu}}
-\frac{2}{P}\frac{\pp^2P}{\pp x^{\mu}\pp x^{\nu}}\right)\right]+V=0
\end{equation}
These equations are equivalent to the Schr\"odinger equation if ${\bf
(C21)}\,\,\psi=\sqrt{P} exp(iS/\hbar)$ with $\gl=(2\hbar)^2$.$\hfill\bs$
\\[3mm]\indent
{\bf REMARK 3.2.}
The SE gives to a probability distribution $\rho=|\psi|^2$ (with suitable
normalization) and to this one can associate an information entropy $S(t)$
(actually configuration information entropy) ${\bf (C22)}\,\,{\mf S}=-\int\rho log(\rho)
d^3x$ which is typically not a conserved quantity.
The rate of change in time of
${\mf S}$ can be readily found by using the continuity equation ${\bf (C23)}\,\,\pp_t\rho=
-\na\cdot(v\rho)$ where $v$ is a current velocity field 
Note here 
(cf. also \cite{p1})
${\bf (C23)}\,\,
\pp{\mf S}/\pp t=-\int\rho_t(1+log(\rho))dx=\int (1+log(\rho))\pp(v\rho)$.
Note that a formal substitution of $v=-u$ in {\bf (C23)} implies the standard free
Browian motion outcome ${\bf (C24)}\,\,d{\mf S}/dt=D\cdot\int [(\na\rho)^2/\rho)d^3x=
D\cdot Tr{\mf F}\geq 0$ - use ${\bf (C25)}\,\,
u=D\na log(\rho)$ with $D=\hbar/2m$) 
and {\bf (C23)} with $\int
(1+log(\rho))\pp(v\rho)=-\int v\rho\pp log(\rho)=-\int
v\rho'\sim\int((\rho')^2/\rho)$ modulo constants involving D etc.   
Recall here ${\mf F}\sim-(2/D^2)\int \rho Qdx=\int dx[(\na \rho)^2/\rho]$ is a functional
form of Fisher information.  A high rate of
information entropy production corresponds to a rapid spreading (flattening down) of
the probablity density.  This delocalization feature is concomitant with the decay in
time property quantifying the time rate at which the far from equilibrium system
approaches its stationary state of equilibrium ${\bf (C26)}\,\,d/dt Tr{\mf F}\leq
0$.$\hfill\bs$ 
\\[3mm]\indent
{\bf REMARK 3.3.}
Comparing now with ${\bf (C1)}\equiv {\bf (C4)}$ or {\bf (C6)} as a Fisher metric we can 
define \eqref{4.9} as a Fisher information metric in the present context.  This should
be positive definite in view of its relation to $(\gD N)^2$ in {\bf (C15)} for example. In
\cite{c1} we sketched many ways in which the quantum potential arises in the derivation
of Schr\"odinger equations.  For $\psi=Rexp(iS/\hbar)$ one has
\bq\label{4.13}
-\frac{\hbar^2}{2m}\frac{R''}{R}\equiv
-\frac{\hbar^2}{2m}\frac{\pp^2\sqrt{\rho}}{\sqrt{\rho}}=-\frac{\hbar^2}{8m}\left[
\frac{2\rho''}{\rho}-\left(\frac{\rho'}{\rho}\right)^2\right]
\end{equation}
in 1-D while in more dimensions we have a form ($\rho\sim P$)
\bq\label{4.14}
Q\sim -2\hbar^2g^{\mu\nu}\left[\frac{1}{P^2}\frac{\pp P}{\pp x^{\mu}}\frac{\pp P}{\pp
x^{\nu}} -\frac{2}{P}\frac{\pp^2P}{\pp x^{\mu}\pp x^{\nu}}\right]
\end{equation}
as in \eqref{4.12} (arising from the Fisher metric I of \eqref{4.9} upon variation in P
in the Lagrangian).  It can also be related to an osmotic velocity field $u=D\na
log(\rho)$ (cf. \cite{g3}) via ${\bf (C26)}\,\,Q=(1/2)u^2+D\na u$ connected to Brownian
motion where D is a diffusion coefficient.  We refer also to \cite{d49,k8,k9,n15,n6}
for other connections to diffusion and statistical mechanics and to \cite{c29,n5}
for origins via a conjectured fractal nature of spacetime (there are also many other
references in \cite{c1}).
$\hfill\bs$

\section{THE SCHR\"ODINGER EQUATION IN WEYL SPACE}
\renewcommand{\theequation}{4.\arabic{equation}}
\setcounter{equation}{0}

A deBroglie-Bohm-Weyl theory has been developed recently by a number of authors (cf.
\cite{c1} for references and a sketch based on the summary article \cite{s3}).  In this
theory one constructs a relativistic framework with quantum matter based in part on
deBroglie - Bohm (dBB) ideas and Weyl geometry.  A Bohmian mass field arises in the
associated Dirac-Weyl theory, corresponding to a quantum mass ${\mf M}$, and the
geometric aspects of the evolving spacetime manifold are related to quantum effects.  A
quantum potential is involved of the form ${\bf (D1)}\,\,{\mf
Q}=(\hbar^2/m^2c^2)(\bx|\Psi|/|\Psi|)$ with
${\bf (D2)}\,\,{\mf M}^2=m^2exp({\mf Q})$.  Evidently probabilistic input to 
the nonrelativistic SE does not apply for relativistic generalizations such as the 
Klein-Gordon (KG) equation and this is eloquently discussed in \cite{n1}.  However
in \cite{s2} one deals with a geometric derivation of the nonrelativistic SE in Weyl
spaces and it turns out that one can relate the standard quantum potential Q to the
Ricci-Weyl scalar curvature of spacetime (see \cite{c1} for details).  The KG equation is
also treated by Santamato in \cite{s2} and the whole matter is analyzed incisevely by
Castro in
\cite{c16}.  Again a relation between the relativistic ${\mf Q}$ and the Weyl-Ricci
curvature exists but without the probabilistic connections.  We remark from \cite{c1},
following \cite{q1,q2,q3,q4}, that one does not expect or want a quantum mechanical
particle to be a free falling trajectory; in the conformal metric the particles do not
follow geodesics of the conformal metric alone.
\\[3mm]\indent 
We refer to \cite{c1,c16,s2} for
philosphy here and to \cite{a12,b3,c1,q1,q2,q3,q4,s1,s3} for Weyl geometry.
In \cite{s2} one begins with a stochastic construction of (averaged) classical type 
Lagrange equations in generalized coordinates for a differetiable manifold M in which a
notion of scalar curvature R is meaningful.  It is then shown that a theory equivalent to
QM (via a SE) can be constructed where the ``quantum force" (arising from a quantum
potential Q) can be related to (or described by) geometric properties of space.  To do
this one assumes that a (quantum) Lagrangian can be constructed in the form
${\bf (D3)}\,\,L(q,\dot{q},t)=L_C(q,\dot{q},t)+\gag(\hbar^2/m)R(q,t)$ where
${\bf (D4)}\,\,\gag=(1/6)(n-2)/ (n-1)$ with $n=dim(M)$ and R is a curvature scalar.
Now for a Riemannian geometry $ds^2=g_{ik}(q)dq^idq^k$ it is standard that
in a transplantation
$q^i\to q^i+\gd q^i$ one has ${\bf (D5)}\,\,\gd A^i=\gG^i_{k\ell}A^{\ell}dq^k$.
Here moreover it is assumed that for $\ell=(g_{ik}A^iA^k)^{1/2}$ one has ${\bf
(D6)}\,\,\gd\ell=\ell\phi_kdq^k$ where the $\phi_k$ are covariant components of an
arbitrary vector of M (Weyl geometry).  For the discussion here we review the
material on Weyl geometry in \cite{s2}.  Thus the actual affine connections
$\gG^i_{k\ell}$ can be found by comparing {\bf (D6)} with $\gd\ell^2=\gd(g_{ik}A^iA^k)$
and using {\bf (D5)}.  A little linear algebra gives then 
\bq\label{6.1}
\gG^i_{k\ell}=-\left\{\begin{array}{c}
i\\
k\,\,\ell\end{array}\right\}+g^{im}(g_{mk}\phi_{\ell}+g_{m\ell}\phi_k-g_{k\ell}\phi_m)
\end{equation}
Thus we may prescribe the metric tensor $g_{ik}$ and $\phi_i$ and determine via \eqref{6.1}
the connection coefficients.  Note that $\gG^i_{k\ell}=\gG^i_{\ell k}$ and for $\phi_i=0$ one
has Riemannian geometry.  Covariant derivatives are defined for contravariant $A^k$ via
${\bf (D7)}\,\,A^k_{,\i}=\pp_iA^k-\gG^{k\ell}A^{\ell}$ and for covariant $A_k$ via
${\bf (D8)}\,\,A_{k,i}=\pp_iA_k+\gG^{\ell}_{ki}A_{\ell}$ (where $S_{,i}=\pp_iS$).  Note 
$g_{ik,\ell}\ne 0$ so covariant differentiation and operations of raising or lowering indices
do not commute.  The curvature tensor $R^i_{k\ell m}$ in Weyl geometry is introduced via
$A^i_{,k,\ell}-A^i_{,\ell,k}=F^i_{mk\ell}A^m$ from which arises the standard formula of
Riemannian geometry ${\bf (D9)}\,\,R^i_{mk\ell}=
-\pp_{\ell}\gG^i_{mk}+\pp_k\gG^i_{m\ell}+\gG^i_{n\ell}\gG^n_{mk}-\gG^i_{nk}\gG^n_{m\ell}$ where
\eqref{6.1} must be used in place of the Riemannian Christoffel symbols.  The tensor
$R^i_{mk\ell}$ obeys the same symmetry relations as the curvature tensor of Riemann geometry as
well as the  Bianchi identity.  The Ricci symmetric tensor $R_{ik}$ and the scalar curvature R
are defined by the same formulas also, viz. $R_{ik}=R^{\ell}_{i\ell k}$ and $R=g^{ik}R_{ik}$. 
For completeness one derives here ${\bf
(D10)}\,\,R=\dot{R}+(n-1)[(n-2)\phi_i\phi^i-2(1/\sqrt{g})\pp_i(\sqrt{g}\phi^i)]$ where 
$\dot{R}$ is the Riemannian curvature built by the Christoffel symbols.  Thus from \eqref{6.1}
one obtains
\bq\label{6.2}
g^{k\ell}\gG^i_{k\ell}=-g^{k\ell}\left\{\begin{array}{c}
i\\
k\,\,\ell\end{array}\right\}-(n-2)\phi^i;\,\,\gG^i_{k\ell}=-\left\{\begin{array}{c}
i\\
k\,\,\ell\end{array}\right\}+n\phi_k
\end{equation}
Since the form of a scalar is independent of the coordinate system used one may compute R in a
geodesic system where the Christoffel symbols and all $\pp_{\ell}g_{ik}$ vanish; then
\eqref{6.1} reduces to ${\bf
(D11)}\,\,\gG^i_{k\ell}=\phi_k\gk^i_{\ell}+\phi_{\ell}\gd^i_k-g_{k\ell}\phi^i$. Hence ${\bf
(D12)}\,\,R=-g^{km}\pp_m\gG^i_{k\ell}+\pp_i(g^{k\ell}\gG^i_{k\ell})+g^{\ell
m}\gG^i_{n\ell}\gG^n_{mi}-g^{m\ell}\gG^i_{n\ell}\gG^n_{m\ell}$.  Further from {\bf (D11)} one
has ${\bf (D13)}\,\,g^{\ell m}\gG^i_{n\ell}\gG^n_{mi}=-(n-2)(\phi_k\phi^k)$ at the point in
consideration.  Putting all this in {\bf (D12)} one arrives at ${\bf (D14)}\,\,
R=\dot{R}+(n-1)(n-2)(\phi_k\phi^k)-2(n-1)\pp_k\phi^k$ which becomes {\bf (D10)} in covariant
form. Now the geometry is to be derived from physical principles so the $\phi_i$ cannot be
arbitrary but must be obtained by the same (averaged) least action principle giving
the motion of the particle.  The minimum is to be evaluated now with respect to
the class of all Weyl geometries having arbitrarily varied gauge vectors but fixed metric
tensor and the only term 
containing the gauge vector is the curvature term.  Then observing that $\gag>0$ when $n\geq 3$
the minimization involves only ${\bf
(D10)}$.  First a little argument shows that
$\hat{\rho}(q,t)=\rho(q,t)/\sqrt{g}$ transforms as a scalar in a coordinate change and this will
be called the scalar probability density of the random motion of the particle.  Starting from
${\bf (D15)}\,\,\pp_t\rho+\pp_i(\rho v^i)=0$ 
a manifestly covariant equation for $\hat{\rho}$ is found to be ${\bf
(D16)}\,\,\pp_t\hat{\rho}+(1/\sqrt{g})\pp_i(\sqrt{g}v^i\hat{\rho})=0$.
Some calculation then yields a minimum for ${\bf
(D17)}\,\,\phi_i(q,t)=-[1/(n-2)]\pp_i[log(\hat{\rho})(q,t)]$.  This shows that the geometric
properties of space are indeed affected by the presence of the particle and in turn the alteration
of geometry acts on the particle through the quantum force $f_i=\gag(\hbar^2/m)\pp_iR$ which
according to {\bf (D10)} depends on the gauge vector and its derivatives.  It is this peculiar
feedback between the geometry of space and the motion of the particle which produces quantum
effects.
\\[3mm]\indent
In this spirit one goes next to a geometrical derivation of the SE.  Thus inserting {\bf (D17)}
into {\bf (D10)} one gets ${\bf
(D18)}\,\,R=\dot{R}+(1/2\gag\sqrt{\hat{\rho}})[1/\sqrt{g})\pp_i(\sqrt{g}g^{ik}\pp_k\sqrt{\rho})]$
where the value {\bf (D4)} for $\gag$ has been used.  On the other hand the HJ equation 
can be written as ${\bf (D19)}\,\,\pp_tS+H_C(q,\na S,t)-\gag(\hbar^2/m)R=0$
where {\bf (D3)} has been used.  When {\bf (D18)} is introduced into {\bf (D19)} the HJ equation
and the continuity equation {\bf (D16)}, with velocity field given by 
${\bf (D20)}\,\,v^i=(\pp H/\pp p_i)(q,\na S,t)$
form a set of two nonlinear PDE which are coupled by the curvature of space.  Therefore self
consistent random motions of the particle (i.e. random motions compatible with {\bf (D12)})
are obtained by solving {\bf (D16)} and {\bf (D19)} simultaneously.  For every pair of solutions
$S(q,t,\hat{\rho}(q,t))$ one gets a possible random motion for the particle whose invariant
probability density is $\hat{\rho}$.  The present approach is so different from traditional QM
that a proof of equivalence is needed and this is only done for Hamiltonians of the form 
${\bf (D21)}\,\,H_C(q,p,t)=(1/2m)g^{ik}(p_i-A_i)(p_k-A_k)+V$
(which is not very restrictive) leading to  
\bq\label{6.3}
\pp_tS+\frac{1}{2m}g^{ik}(\pp_iS-A_i)(\pp_kS-A_k)+V-\gag\frac{\hbar^2}{m}R=0
\end{equation}
(R in {\bf (D18)}).  The continuity
equation {\bf (D16)} is ${\bf
(D22)}\,\,\pp_t\hat{\rho}+(1/m\sqrt{g})\pp_i[\hat{\rho}\sqrt{g}g^{ik}(\pp_kS-A_k)]=0$.
Owing to {\bf (D18)} \eqref{6.3} and {\bf (D22)} form a set of two nonlinear PDE which must be
solved for the unknown functions S and $\hat{\rho}$.  Then a straightforward calculations shows
that, setting ${\bf (D23)}\,\,\psi(q,t)=\sqrt{\hat{\rho}(q,t)}exp](i/\hbar)S(q,t)]$, the
quantity
$\psi$ obeys a linear PDE (corrected from \cite{s2})
\bq\label{6.4}
i\hbar\pp_t\psi=\frac{1}{2m}\left\{\left[\frac{i\hbar\pp_i\sqrt{g}}{\sqrt{g}}+A_i
\right]g^{ik}(i\hbar\pp_k+A_k)\right\}\psi+\left[V-\gag\frac{\hbar^2}{m}\dot{R}\right]\psi=0
\end{equation}
where only the Riemannian curvature $\dot{R}$ is present (any explicit reference to the gauge
vector $\phi_i$ having disappeared).  \eqref{6.4} is of course the SE in curvilinear coordinates
whose invariance under point transformations is well known.  Moreover {\bf (D23)} shows that
$|\psi|^2=\hat{\rho}(q,t)$ is the invariant probability density of finding the particle in the
volume element $d^nq$ at time t.  Then following Nelson's arguments that the SE together with
the density formula contains QM the present theory is physically equivalent to traditional
nonrelativistic QM.  
\\[3mm]\indent
{\bf REMARK 4.1.}
We recall (cf. \cite{c1}) that in the nonrelativistic context the quantum potential has the form
$Q=-(\hbar^2/2m)(\pp^2\sqrt{\rho}/\sqrt{\rho})\,\,(\rho\sim\hat{\rho}$ here) and in more
dimensions this corresponds to $Q=-(\hbar^2/2m)(\gD\sqrt{\rho}/\sqrt{\rho})$.  The continuity
equation in {\bf (D22)} corresponds to
$\pp_t\rho+(1/m\sqrt{g})\pp_i[\rho\sqrt{g}g^{ik}(\pp_kS)]=0$
($\rho\sim\hat{\rho}$ here).  For $A_k=0$
\eqref{6.3} becomes ${\bf (D24)}\,\,\pp_tS+(1/2m)g^{ik}\pp_iS\pp_kS+V-\gag(\hbar^2/m)R=0$.
This leads to an identification ${\bf (D25)}\,\,Q\sim-\gag(\hbar^2/m)R$ where R is the Ricci
scalar in the Weyl geometry (related to the Riemannian curvature built on standard Christoffel
symbols via {\bf (D10)}).  Here $\gag=(1/6)[(n-2)(n-2)]$ as in {\bf (D4)} which for $n=3$
becomes $\gag=1/12$; further by {\bf (D17)} the Weyl field $\phi_i=-\pp_i log(\rho)$.
Consequently
for the SE \eqref{6.4} in Weyl space the quantum potential is $Q=-(\hbar^2/12m)R$ where R
is the Weyl-Ricci scalar curvature.  For Riemannian flat space $\dot{R}=0$ this becomes
via {\bf (D18)}
\bq\label{6.5}
R=\frac{1}{2\gag\sqrt{\rho}}\pp_ig^{ik}\pp_k\sqrt{\rho}\sim\frac{1}{2\gag}
\frac{\gD\sqrt{\rho}}{\sqrt{\rho}}\Rightarrow
Q=-\frac{\hbar^2}{2m}\frac{\gD\sqrt{\rho}}{\sqrt{\rho}}
\end{equation}
as is should and the SE \eqref{6.4} reduces to the standard SE $i\hbar\pp_t\psi=
-(\hbar^2/2m)\gD\psi+V\psi$ ($A_k=0$).
$\hfill\bs$
\\[3mm]\indent
{\bf REMARK 4.2.}
The formulation above from \cite{s2} is also developed for a derivation of the
Klein-Gordon (KG) equation via an average action principle with the restrictions of Weyl
geometry released.  The spacetime geometry was then obtained from the action principle to
obtain Weyl connections with a gauge field $\phi_{\mu}$.  The Riemann scalar curvature 
$\dot{R}$ is then related to the Weyl scalar curvature R via an equation ${\bf (D26)}\,\,
R=\dot{R}-3[(1/2)g^{\mu\nu}\phi_{\mu}\phi_{\nu}+(1/\sqrt{-g})\pp_{\mu}(\sqrt{-g}g^{\mu\nu}
\phi_{\nu}]$.  Explicit reference to the underlying Weyl structure disappears in the
resulting SE (as in \eqref{6.4}).  The HJ equation in \cite{s2} has then the form (for
$A_{\mu}=0$ and $V=0$) ${\bf (D27)}\,\,g^{\mu\nu}\pp_{\mu}S\pp_{\nu}S=m^2-(R/6)$ so in
some sense (recall here $\hbar=c=1$) ${\bf (D28)}\,\,m^2-(R/6)\sim {\mf M}^2$ (via
arguments in \cite{s2} - cf. also \cite{c1}) where ${\mf M}^2=m^2exp({\mf Q})$ and
${\mf Q}=(\hbar^2/m^2c^2)(\bx\sqrt{\rho}/\sqrt{\rho})\sim
(\bx\sqrt{\rho}/m^2\sqrt{\rho})$ (for signature $(-,+,+,+)$).  Thus for $exp({\mf Q})\sim
1+{\mf Q}$ one has ${\bf (D29)}\,\,m^2-(R/6)\sim m^2(1+{\mf Q})\Rightarrow (R/6)\sim
-{\mf Q}m^2\sim -(\bx\sqrt{\rho}/\sqrt{\rho})$.  This agrees also with \cite{c16} where
the whole matter is analyzed incisively (and we recall the remarks at the beginning of
Section 4).  In this situation the probabilistic aspects (if any) are hidden and we refer
to \cite{n1} for discussion of this point.
$\hfill\bs$
\\[3mm]\indent
{\bf REMARK 4.3.}
For $\dot{R}=0$ one has as in Remark 4.1 $Q\sim(\gag\hbar^2/m)R$ where $\gag=1/12$ with
${\bf (D30)}\,\,R=(1/2\gag\sqrt{\rho})\pp_ig^{ik}\pp_k\sqrt{\rho}=(1/2\gag\sqrt{\rho})g^{ik}
\pp_i\pp_k\sqrt{\rho}$ (since $g^{ik}$ can be taken to be constant - cf. \cite{a21}).
Then writing out \eqref{6.5} we have 
\bq\label{6.6}
Q=-\frac{\hbar^2}{2m}\frac{1}{\sqrt{\rho}}g^{ik}\pp_i\pp_k\sqrt{\rho}=\frac{\hbar^2g^{ik}}{8m}
\left(\frac{2\pp_i\pp_k\rho}{\rho}-\frac{\pp_i\rho\pp_k\rho}{\rho^2}\right)
\end{equation}
correponding to \eqref{4.14}.  Thus Q and consequently $R=-(m/\gag\hbar^2)Q$ arise from
variation of the Fisher metric I of \eqref{4.9} in $P\sim\rho$.  Noting that (as in Remark
3.2) integrals of the form $\int \pp_i\pp_k\rho d^3x$ could be expected to vanish for
distributions $\rho$ decreasing rapidly with their derivatives at $\infty$ we could say now
that ${\bf (D31)}\,\,\int \rho Qd^3x\sim -(\hbar^2
g^{ik}/8m)\int[\pp_i\rho\pp_k\rho/\rho)d^3x=-(\hbar^2/8m)I$ via \eqref{4.9}.  This says that
$(\gag=1/12$) ${\bf (D32)}\,\,I\sim-(\hbar^2/8m)\int\rho[-(\gag\hbar^2/m)R]d^3x=(\hbar^4/96
m^2)\int\rho Rd^3x$ and presents an explicit connection between the Fisher information metric
and the Weyl-Ricci scalar curvature R (for Riemann flat spaces).$\hfill\bs$
\\[3mm]\indent
{\bf ACKNOWLEDGEMENT.}
To my wonderful wife Denise.

\end{document}